\documentclass[prb,twocolumn]{revtex4-1}

\usepackage{amsmath}  
\usepackage{amsfonts} 
\usepackage{graphicx} 

\begin{document}


\title{Resolving Tensions Surrounding Massive Pulleys}

\author{D. J. Durian, J. Kroll and E. J. Mele}
\email{mele@physics.upenn.edu}
\altaffiliation[permanent address:] {\\ Department of Physics and Astronomy \\ David Rittenhouse Laboratory \\ University of Pennsylvania \\ Philadelphia PA 19104 \\ USA \\ }
\affiliation{Department of Physics and Astronomy \\ David Rittenhouse Laboratory \\ University of Pennsylvania \\ Philadelphia PA 19104  USA }


\date{\today}

\begin{abstract}
The distribution of string tension on the contact line between an ideal string and a massive pulley is a frequently-discussed but incompletely-posed problem that confronts students in introductory mechanics. We highlight ambiguities in the usual presentation of this problem by the massive Atwood's machine and discuss two compact resolutions that treat situations where the pulley or the string elastically deform.  We propose experiments that can be developed in an intermediate laboratory to determine the tension profile.
\end{abstract}

\maketitle 

\section{Introduction} 
Talking honestly to introductory students about friction is always difficult. The conventional Amontons-Coulomb formulation of friction \cite{friction,friction2,YFSMf} does not prescribe a force ``law" but is more in the spirit of a force ``rule". Furthermore the rule for static friction $f \le \mu_s N$ (with coefficient of static friction $\mu_s$  and normal force $N$) does not specify the force at all but rather gives a stability limit for enforcing a no-slip condition.  Comparisons of published lists of coefficients of friction turn up significant discrepancies between them for many interfaces. This is for a good reason: the measurements are hard to reproduce and depend on the preparation, quality and cleanliness of contacted surfaces, as well as loading history, none of which are controlled or even controllable in practical experimental settings. Classroom demonstrations of frictional forces are notoriously resistant to reproducible quantitative analysis.  When it comes to friction, introductory students expect to be handed an algorithm but instead are given only conditional advice.

Contributing to the confusion is the problematic issue of understanding {\it distributed} frictional force along the contact line between a string and a massive pulley.  A conventional situation is encountered in the massive Atwood's machine \cite{YFSM,Greenslade} (Figure~\ref{atwoods}) where one asserts that when friction enforces a no slip condition along the contact line between a massless inextensible string and a massive pulley the tension must vary as a function of position on the string.  Here students are asked to abandon their trusted beliefs that the masslessness of the string would guarantee that the tension is {\it uniform} even in this nonequilibrium setting \cite{massless}. This understanding needs to be discarded in favor of a new principle stating that actually the tension must vary, but in a way that we can't describe except to determine its boundary values where the string meets and leaves the pulley \cite{Krause,Lemos}.  Usually this issue arises at the very time in the course when the rules of rigid body motion appear to be systematically evading other seemingly immutable laws of point particle dynamics, lending a certain atmosphere of incredulity to the entire discussion. Don't even bring up the subject of rolling friction \cite{rolling} or the distinction between ``dry" and ``wet'' friction \cite{dryandwet}, etc.

The purpose of this article is to revisit this problem and discuss physical resolutions to the massless string-massive pulley dilemma. The resolutions are interesting and show students something important. This problem is confusing because, as usually formulated, it admits a family of possible solutions rather than a single prescribed solution. This degeneracy can be lifted by extending the problem to describe the elastic compliances of the string and pulley. Ordinarily one would expect this refinement to take the discussion well outside the domain of any course in elementary mechanics \cite{hyperstatic}. However these issues {\it can} be meaningfully discussed even in an undergraduate course. Here we present two competing resolutions both of which are pedagogically useful and should be accessible to intermediate students. One of these makes contact with the outcome that one might have intuitively guessed at the outset though without any justification.  The other finds some qualitative support in experimental measurements of the tension profile in a drive belt. Comparison of these two models allows a useful discussion of principles that {\it could} determine the friction distribution in the string. In the last section of this paper we suggest an experiment that can be designed and built to provide a refined test of these principles.

\section{Massive Atwood's machines with and without Slipping}

\begin{figure}[h!]
\centering
\includegraphics[width=\columnwidth]{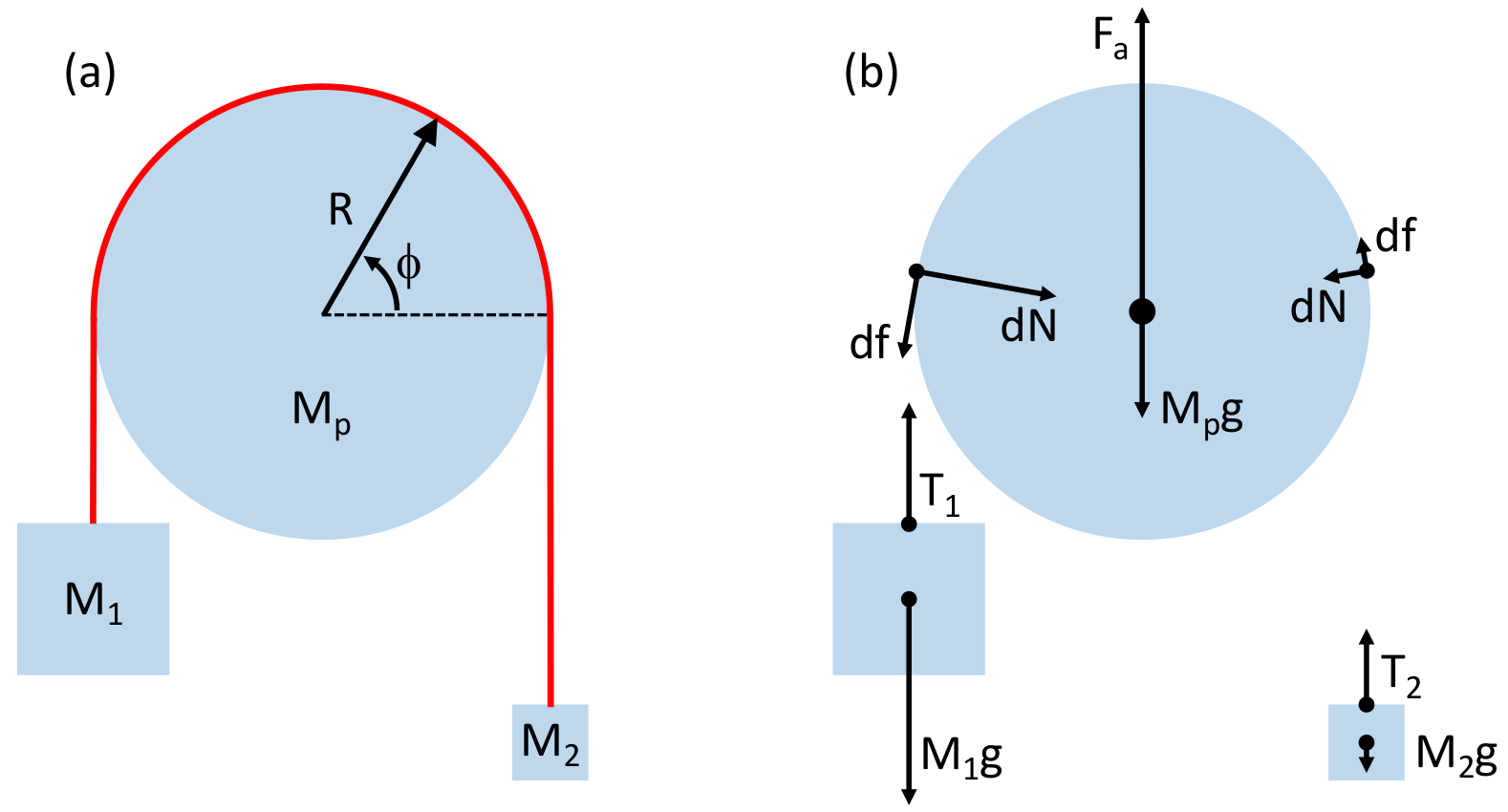}
\caption{(a) Atwood's machine consists of a frictionless pulley of mass $M_p$ and two masses $M_1>M_2$ connected a massless inextensible string (shown in red).  (b) The masses are subject to gravity and tension forces, while the pulley experiences distributed normal ($dN$) and frictional ($df$) contact forces from the string as well as gravity $M_p g$ and a force from the axle $F_a$.}
\label{atwoods}
\end{figure}

For completeness we collect a few familiar results in the solution of the massive Atwood's machine with the no slip condition (Fig.~\ref{atwoods}a). For definiteness consider the case where $M_1 > M_2$ so the pulley rotates counterclockwise as the masses fall and rise.   The standard calculation \cite{YFSM} for a pulley with moment of inertia $M_P R^2/2$ with the no slip condition $\alpha = - a_1/R$ gives for the acceleration of $M_1$
\begin{eqnarray}
a_1 = \frac{(M_2 - M_1)\, g}{M_1 + M_2 + \frac{1}{2} M_P} <0 \,\, {\rm (it \, falls)} \nonumber
\end{eqnarray}
Therefore the tension difference between the two ends is
\begin{eqnarray}
T_1 - T_2 = \frac{1}{2} \frac{M_p(M_1 - M_2)}{M_1 + M_2 + \frac{1}{2} M_p} \,  g \nonumber
\end{eqnarray}
and the explicit forms for $T_1$ and $T_2$ are
\begin{eqnarray}
T_1 &=& \frac{2 M_1 M_2 + \frac{1}{2} M_1 M_p}{M_1 + M_2 + \frac{1}{2} M_p} \, g \nonumber\\
T_2 &=& \frac{2 M_1 M_2 + \frac{1}{2} M_2 M_p}{M_1 + M_2 + \frac{1}{2} M_p} \, g \nonumber
\end{eqnarray}
so that $T_1$ is the larger and the two tensions revert to the same value if $M_p \rightarrow 0$.  It is worth noting that if there is no slipping these expressions can be deduced by energy conservation without any knowledge of the tension profile or the actual string-pulley normal and friction contact forces (Fig.~\ref{atwoods}b).

\subsection{String-Pulley Contact Forces}

\begin{figure}[ht!]
\centering
\includegraphics[width=1.6in]{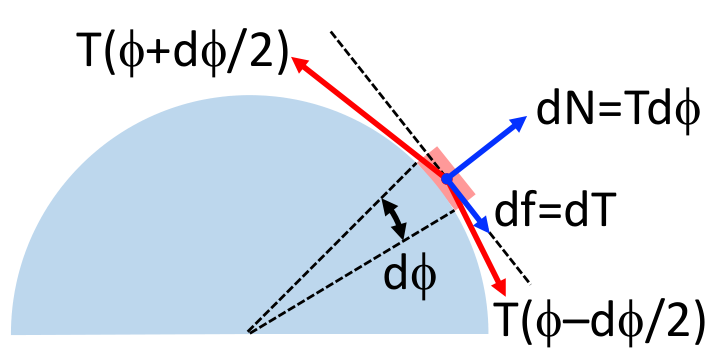}
\caption{A curved differential element of the string is pulled from each end by the tension in adjacent elements.  If the string is massless, this is balanced by a normal force $dN$ and a friction force $df$ exerted by the pulley with indicated values dictated by the tension profile, $T(\phi)$.
\label{rc2}}
\end{figure}

The transition from $T_1$ to $T_2$ is determined by the string-pulley interaction along the contact line.  A useful free-body diagram for a line element of the string is shown in Figure~\ref{rc2}. Since the string is massless we get for the force along the local tangent line
\begin{eqnarray}
T(\phi- d \phi/2) \cos (d \phi/2)  + df - T(\phi + d \phi/2) \cos (d \phi/2)  &=& 0 \nonumber
\end{eqnarray}
so this gives
\begin{eqnarray}
\frac{d T}{d \phi}  = \frac{df}{d \phi}.
\end{eqnarray}
Note that the total torque exerted by friction from the string on the pulley is thus always $\int R df = \int R dT = R(T_1-T_2)$ for any tension distribution.  This result is implicit (though it is usually concealed) in the diagram one conventionally draws for the external forces acting on the pulley.  Next, projecting similarly along the normal direction gives
\begin{eqnarray}
dN - 2 T(\phi) \sin(d \phi/2) = 0 \nonumber
\end{eqnarray}
so
\begin{eqnarray}
\frac{dN}{d \phi} = T(\phi) \nonumber
\end{eqnarray}
Note that the stability limit for static friction is
\begin{eqnarray}
|df| \leq \mu \, dN
\end{eqnarray}
where $\mu$ is the coefficient of static friction. Combining Eqs.~(1) and (2) tells us that the stability condition puts a bound on how fast the tension can vary as a function of angle.
\begin{eqnarray}
\left| \frac{d T}{d \phi} \right|  \le  \mu T(\phi)\Rightarrow
\left| \frac{d \log T}{d \phi} \right|  \le  \mu  \nonumber\\
\end{eqnarray}
If this criterion fails, then the string slips and the tension must satisfy $| dT/d\phi | = \mu_k T(\phi)$ where $\mu_k$ is the coefficient of kinetic friction.  These arguments~\cite{Krause,Lemos} may be generalized for strings that are massive and extensible~\cite{Bechtel}.

\subsection{Broken Atwood's Machine}

The physics of Figs.~\ref{atwoods}-\ref{rc2} and Eqs.~(1-3) for local string-pulley contact forces can be brought to bear at an elementary pedagogical level for a ``broken" Atwood's machine where the pulley is not allowed to rotate.  For example, the largest hanging mass $M_1$ that can be supported in static equilibrium by a smaller hanging mass $M_2$ is such that Eq.~(3) is an equality and the tension increases exponentially from $T_2=M_2g$ to $T_1=M_1g$ as $T(\phi)=T_2\exp(\mu \phi)$.  The largest supportable mass is thus exponentially larger, $M_1=M_2\exp(\mu \pi)$, and there is no ambiguity in the tension profile.  Here the fixed pulley serves as a capstan \cite{Hazelton,KK}, which can support a force that is larger by a factor of up to $\exp(\mu \Omega)$ according to the total winding angle $\Omega$.  As another example, if $M_1$ is too large then it will accelerate down at rate $a$ given by solution of $M_1g-T_1=M_1a$ and $T_2-M_2g=M_2a$ where $T_1=T_2 \exp(\mu_k\pi)$ and $\mu_k$ is the coefficient of sliding friction.  The result is $a=[M_1-M_2\exp(\mu_k\pi)]/[M_1+M_2\exp(\mu_k\pi)]$ and the tension profile is again exponential along the entire contact arc, $T(\phi)=T_2\exp(\mu_k\phi)$ with $T_2=2M_1M_2g/[M_1+M_2\exp(\mu_k\pi)]$.

\section{Crossover Solutions}

The solution to Equation 3 gives the limits of stability for the no slip constraint on the contact line.  A crucial point is that it is a {\it first order differential equation} so that the stability limit everywhere is specified by only one constant of integration that could be imposed independently at either end. This feature is exploited in applications of friction to exponential force amplification in the capstan \cite{Hazelton,KK}.  However this one-sided character also exposes  an unavoidable frustration for the Atwood's machine: imposing the boundary condition on a single end of the pulley can (and generally does) invite a violation of the stability conditions elsewhere along the contact line when an additional boundary condition is imposed.

\begin{figure}[ht!]
\centering
\includegraphics[width=\columnwidth]{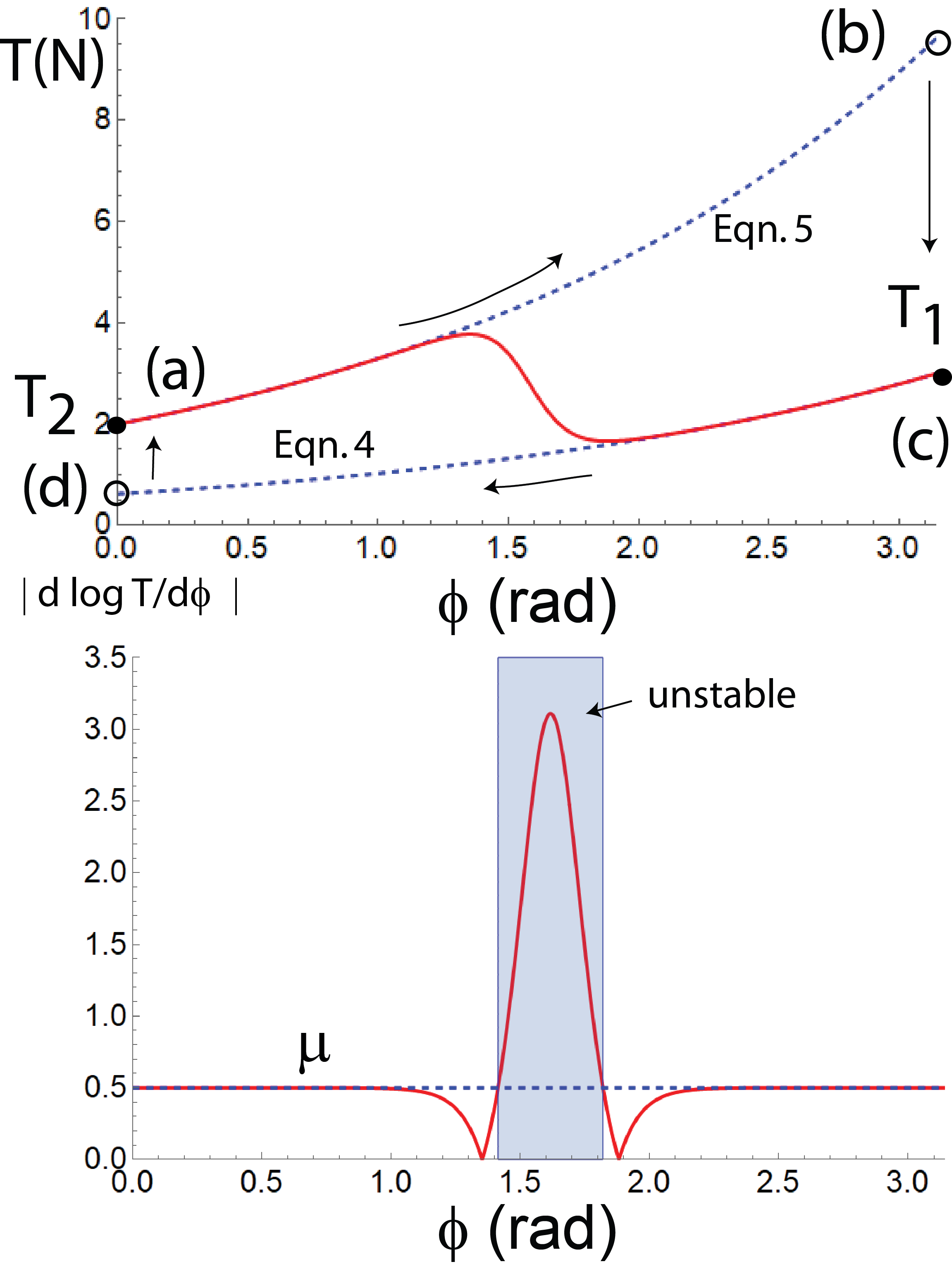}
\caption{(top) Tension profiles in marginally stabile and unstable tension profiles (here $T_1 = 3 \, {\rm N}, T_2 = 2 \, {\rm N}, \mu_s = 0.5$). Imposing the boundary condition $T(0) = T_2$ defines a marginally stable solution (dashed line: a-b from Eqn. 4) that does not match the target state $T_1$. Similarly the boundary condition $T(\pi) = T_1$ gives a marginally stable solution (dashed line: c-d Eqn. 5) that does not match the boundary value $T(0) = T_2$.  For a critical tension profile at both ends one requires a   crossover solution. One such crossover solution  shown in red interpolates smoothly between these limits but it is unstable in the middle of the string. (bottom)  The stability condition $|d \log T/d \phi|< \mu$ (Eqn. 3) is violated by the crossover function within the region indicated by the shaded box.
\label{rc3}}
\end{figure}

As an illustration, let's suppose that the string is marginally stable at the left side of Figure 1 at $\phi = \pi$ (with tension $T_1$). Then following the stability condition along the string we find (Fig. 3)
\begin{eqnarray}
T(\phi) \le  \frac{(2 M_1 M_2 + \frac{1}{2} M_1 M_p) \, g }{M_1 + M_2 + \frac{1}{2} M_p} \,\, \exp(\mu \phi - \mu \pi) \nonumber
\end{eqnarray}
which  at the right hand string  $\phi = 0$ evaluates to
\begin{eqnarray}
T(\phi=0) &=& \frac{(2 M_1 M_2 + \frac{1}{2} M_1 M_p) \, g }{M_1 + M_2 + \frac{1}{2} M_p} \,\,  \exp( - \mu \pi) \nonumber\\
&<& T_2
\end{eqnarray}
This value is exceeded by the {\it required} magnitude of the tension $T_2$ in the right hand string. Conversely if the segment on the right  hand side is marginally stable with tension $T_2$ (Fig. 3) one has instead
\begin{eqnarray}
T(\phi)  \le  \frac{(2 M_1 M_2 + \frac{1}{2} M_2 M_p) \, g }{M_1 + M_2 + \frac{1}{2} M_p} \exp(\mu \phi) \nonumber
\end{eqnarray}
whose value is consistent with the (smaller) required tension on the left $T_1$, i.e.
\begin{eqnarray}
T(\pi) = T_1 \le \frac{(2 M_1 M_2 + \frac{1}{2} M_2 M_p) \, g }{M_1 + M_2 + \frac{1}{2} M_p} \,\, \exp(\pi \mu)
\end{eqnarray}
However the stability condition at intermediate angles is more subtle: it relates the value {\it and} the slope of $T(\phi)$ at {\it all} intermediate angles.   Thus if we start from a marginally on the right solution one actually needs to compare
\begin{eqnarray}
\left| \frac{dT}{d \phi} \right| \,\, {\rm with} \,\,\mu T(\phi) \nonumber
\end{eqnarray}
or more precisely
\begin{eqnarray}
\left| \frac{d \log T}{d \phi} \right| \le \mu \nonumber
\end{eqnarray}
This shows that any crossover between marginally stable profiles on the right $(T_2)$ and on the left end $(T_1)$ must {\it also} be sufficiently soft to satisfy this stability condition {\it everywhere} in the interior.  As an example Figure~\ref{rc3} illustrates a plausible smooth differentiable crossover profile that satisfies the stability conditions near both endpoints but fails to satisfy the ``sufficiently smooth" condition in the interior.

\begin{figure}[ht!]
\centering
\includegraphics[width=\columnwidth]{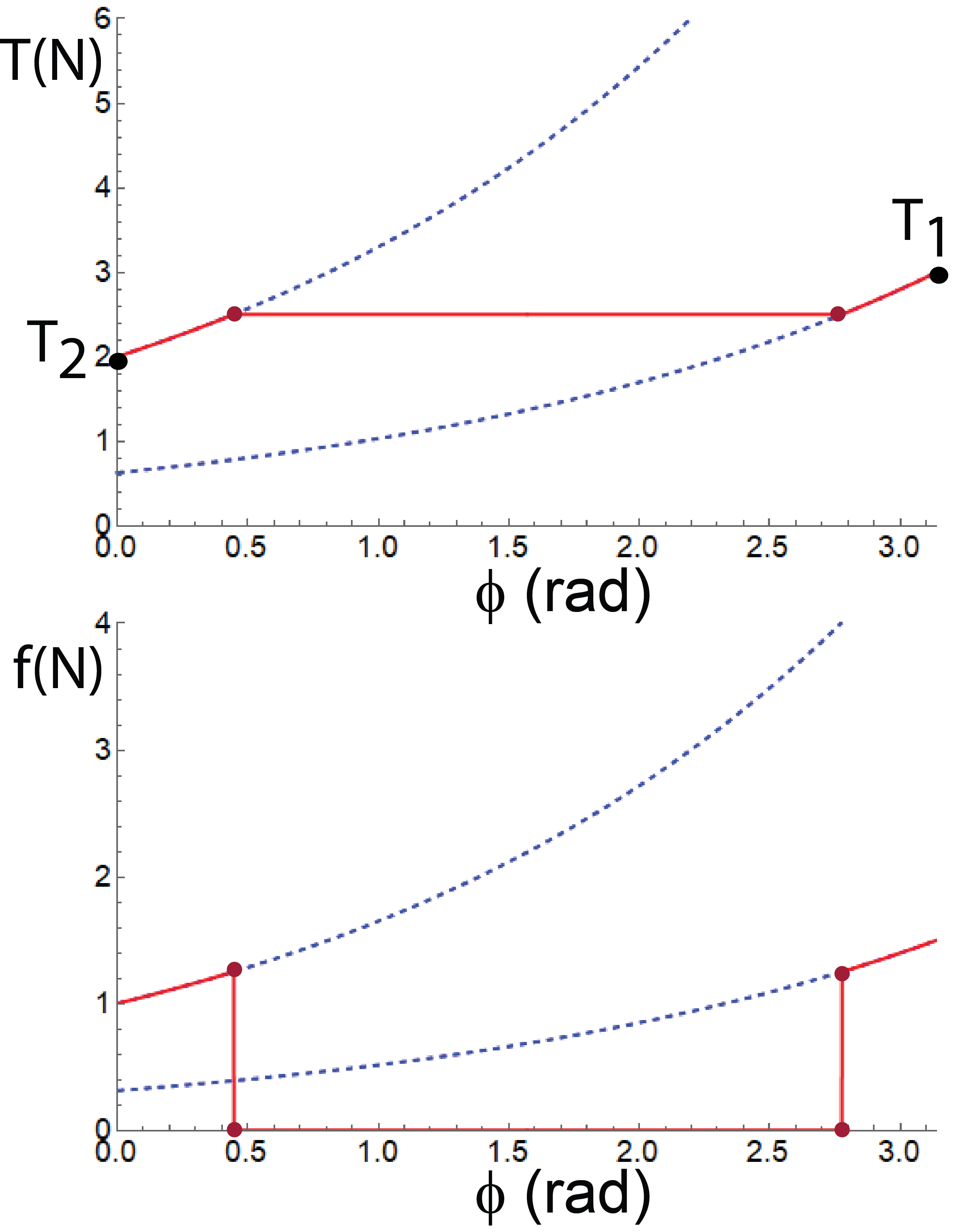}
\caption{(top) A constant tension (frictionless) domain separates regions described by the asymptotic marginally stable states that meet the boundary conditions at $\phi = 0$ and $\phi=\pi$ (here $T_1 = 3 \, {\rm N}, T_2 = 2 \, {\rm N}, \mu_s = 0.5$). (bottom) This piecewise continuous friction profile is nulled in the middle region and satisfies the stability condition everywhere with a piecewise continuous tension profile. Dashed lines in both panels are the marginally stable tension profiles matched to the boundary conditions at the two ends as illustrated in Figure 3.
\label{rc3p2}}
\end{figure}

There are a family of stable crossover solutions where the tension connects the boundary values at both ends  and without violating the slope condition anywhere in the interior. Any such crossover solution cannot be ``too steep"  since its slope at any angle determines the local friction. However this is a very loose condition and there remains a large family of admissible tension profiles.  One important profile that satisfies all the conditions is the ``frictionless crossover" solution shown in Figure~\ref{rc3p2}. Here the system actually phase separates into three distinct regions: the disconnected bounding regions are marginally stable and satisfy the boundary values separately at both endpoints separated by a third domain where the tension is {\it constant}. Noting that $dT/d \phi$ determines the local friction, in this middle region there is actually {\it no} friction acting along the contact line. The location and width of this frictionless crossover domain is of course completely arbitrary. This system is phase-separated so all the torque is provided only in the bounding regions and the central domain floats, comoving with the rotating pulley.

Recognizing the possibility of heterogeneous phase-separated solutions it is  immediately clear that one can generalize the argument to produce a large family of admissible distributed tension profiles. This is an inevitable consequence of adopting an incomplete specification of a force law  which could prescribe the distributed friction along the contact line. This type of ambiguity arises even in equilibrium problems and is similar in spirit to the dilemma of trying to identify the distribution of normal forces that  support a four-legged stool  \cite{hyperstatic}.

\section{Compliant Atwood's Machines}

\smallskip
One resolution of this degeneracy comes from a consideration of the elastic compliance of the system. Below a critical loading state a tension profile that minimizes the stored elastic energy turns out to be a single domain solution with a constant slope $dT/d \phi$ in the interior. Interestingly, when confronted with the mismatch $T_1 - T_2$ required for the rigid body rotation of the pulley, one might naively guess that a linear interpolation between the endpoints would occur. We now demonstrate how this guess is realized with a quantitative analysis. Consideration of this family of solutions identifies a two-domain solution in the approach to a critical state where the no slip condition is  ultimately violated.

\smallskip
Returning to the free body diagram for a massless rope we have
 \begin{eqnarray}
 df = \frac{dT}{d \phi} \, d \phi \nonumber
 \end{eqnarray}
 The application of this force on the boundary of an elastically compliant medium slightly changes its shape and produces a stored elastic energy. This can be calculated by considering the work done by the string as elements of the boundary undergo differential displacements. These displacements are in turn {\it induced} by forces from the string, and for an elastic medium this energy can be calculated by summing over pairwise products of the differential forces applied on the boundary.  For an elastic medium this coupling is both nonlocal and long range. In this way, regarding the string and the pulley as elastically compliant media  the stored elastic energy can be represented by a nonlocal functional of the $T(\phi)$ profile
 \begin{eqnarray}
 U_{\rm elastic} = \frac{1}{2} \, \int_0^\pi \, d \phi \, d \phi' \,\, T'(\phi) \,  G(\phi - \phi') \, T'(\phi')
 \end{eqnarray}
For the cylindrical geometry relevant to the pulley we observe that this interaction kernel $G(u)$ must be a periodic and even function of the difference angular coordinate $u$
 \begin{eqnarray}
 G(u) = g_0 + g_1 \cos (u) \nonumber + \ldots.
 \end{eqnarray}
Inserting this into the expression for $U_{\rm elastic}$ one finds
 \begin{eqnarray}
 U_{\rm elastic} = \frac{1}{2} g_0 \left| t_0 \right|^2 + \frac{1}{2} g_1 \left| t_1 \right|^2  + \ldots
 \end{eqnarray}
 where
 \begin{eqnarray}
 t_m = \int_0^\pi \, d \phi \,\, \frac{dT}{d \phi} \, e^{-im \phi}. \nonumber
 \end{eqnarray}
 Note  that the zeroth order term is
 \begin{eqnarray}
 t_0 = T_1 - T_2 \nonumber
 \end{eqnarray}
and this is constrained by the macroscopic equation of motion. But $t_1$ and the higher moments are not similarly constrained, and since their contribution to the elastic energy are independent and positive definite, we immediately conclude that the optimal solution has $t_1$ (and all the higher moments) equal to zero. The one remaining nonvanishing zeroth moment is in fact the constant slope solution. It is the preferred distribution in the sense that it minimizes the stored elastic energy.  The interpretation of this result is that to minimize the energy stored in an elastically compliant medium there is a tendency to make the distribution of friction on the contact line as smooth as possible.

 \smallskip
Strictly speaking if the string were inextensible string and the pulley perfectly rigid,  then $g_1$ (and all the higher $g_m$'s) are zero. In that case there is no deformation, no stored elastic energy and then  all the solutions discussed above would have the same energy, {\it i.e.}\ they are all equally likely. To understand the tension profile one needs to avoid unphysical ideal-string-and-rigid-pulley constraints. It is also interesting that this ``distributed tension" solution is  preferred over even the ``frictionless domain" solution. This is because the dynamics requires that there must be friction {\it somewhere} along the contact line, and using our expression for $U_{\rm elastic}$ we see that it actually costs a larger elastic energy to confine this frictional coupling to the bounding segments (or in any inhomogeneous state)  than to just distribute it uniformly. Since $f = dT/d \phi$ this solution minimizes the spatial variations of the friction (and  hence the stored elastic energy) by it spreading out uniformly over the entire contact line.

\begin{figure}[ht!]
\centering
\includegraphics[width=\columnwidth]{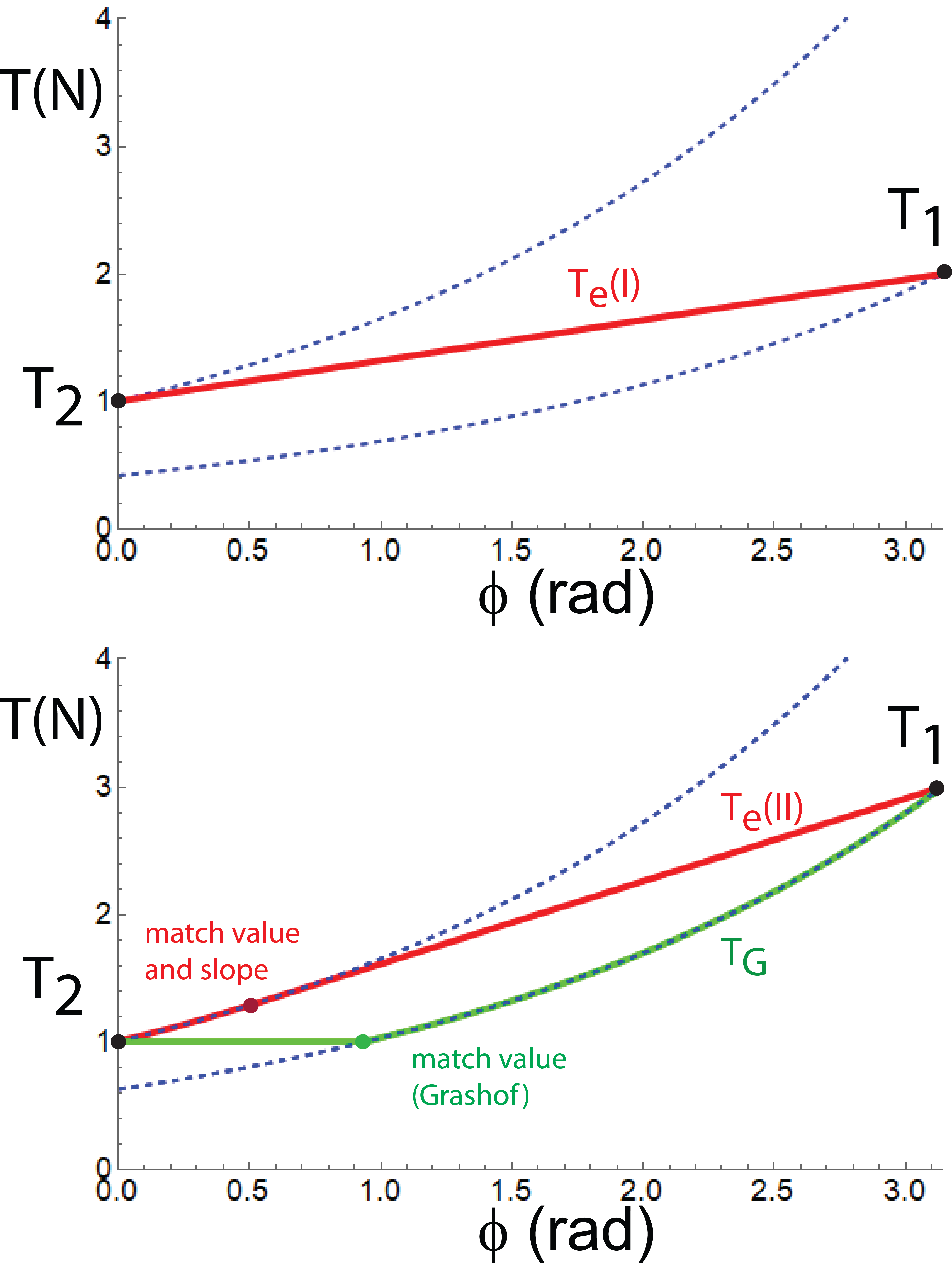}
\caption{ The dashed lines give the critical tension profiles on the verge of slipping with boundary conditions $T_1$ and $T_2$. (top) The elastically optimized tension profile $T_e(I)$  bounded by the critical forces is a single domain that linearly interpolates between the two boundary values $T_1$ and $T_2$. (bottom) Two stable phase separated tension profiles.  $T_e(II)$ gives the piecewise continuous elastically minimized solution where the region of {\it lower} tension is in the critical state. $T_G$ gives the  Grashof two-domain model where the region of {\it higher} tension is in the critical state.
\label{critical}}
\end{figure}

\smallskip
Armed with this information we can also address the seldom-answered question of when and how the no-slip condition fails for the massive Atwood's machine. The linear tension profile gives
\begin{eqnarray}
T(\phi) = T_2 + (T_1 - T_2) \frac{\phi}{\pi} \nonumber
\end{eqnarray}
and a uniform friction distribution
\begin{eqnarray}
f = \frac{dT}{d \phi} = \frac{T_1 - T_2}{\pi} \nonumber
\end{eqnarray}
Since
\begin{eqnarray}
\frac{T_1 - T_2}{\pi} \le \mu \left[T_2 + (T_1 - T_2) \, \frac{\phi}{\pi} \right] \nonumber
\end{eqnarray}
we have
\begin{eqnarray}
\frac{T_1 - T_2}{T_2 + (T_1 - T_2) \, \phi/\pi} \le \pi \mu \nonumber
\end{eqnarray}
Thus the region of string closest to the smaller mass is more susceptible to slipping which occurs at the critical tension ratio
\begin{eqnarray}
\frac{T_1}{T_2}  = 1 + \mu \pi \nonumber
\end{eqnarray}
For $\mu \sim 0.5$ this is a critical tension ratio $\sim 2.6$. Above this value the region of the string under higher tension can still be locally stable. In this solution the system phase separates into a stable domain with a linear tension profile shown in Figure~\ref{critical}. The segment closer to the larger mass has a {\it constant} distributed friction) with a tension profile that matches both the value and slope of the profile in a second marginally stable domain with {\it exponentially growing} friction. With increasing $T_1/T_2$ the latter mobilized region expands to fill the entire contact line and the system reverts to sliding motion when the ratio of boundary tensions exceeds the capstan bound \cite{Hazelton,KK}
\begin{eqnarray}
\frac{T_1}{T_2} > e^{\, \mu \pi} \nonumber
\end{eqnarray}
For $\mu = 0.5$ the critical force ratio  for slipping is thus $\sim 4.8$. Interestingly, this mode of instability describes a collective avalanche breakdown which occurs simultaneously over the entire contact line even though the friction necessarily varies as a function of angular position.

\medskip
\section{Kinetically Limited Solutions}

\smallskip
Commonly, frictional forces on the contact line can be history dependent.  If one allows for the possibility of slip at the interface, relaxation to an elastically optimized tension profile may be practically impossible, since reaching this state may require exploring inaccessible tension configurations. We refer to these solutions as being kinetically  limited.  The possibility of these solutions requires one to confront the existence of possibly truly indeterminate force distributions in the dynamics.  Although this is a daunting possibility, a compact description of this type of kinetically limited behavior is found in the Grashof model developed in the nineteenth century \cite{Grashof, Johnson}, and tested experimentally in recent years \cite{KMN, DPT} for the transmission of torque from a rigid {\it driving} pulley to a rigid {\it driven} pulley via elastic belt. Remarkably this model gives a resolution that is the {\it converse} to the elastically minimized solutions discussed in Section IV.  For the {\it driven} pulley, the Grashof model asserts that the belt-pulley contact  always separates into two regions.  Upon entry of the belt onto the pulley from the low-tension side, one posits a ``stick arc" where the tension is constant and hence no frictional torque is exerted.  On the other hand on the high-tension exit side, there is a ``slip arc" where the string stretches slightly and exerts a sliding friction responsible for the torque.  In this second domain the tension rises exponentially from $T_1$ to $T_2$ according to the usual Coulomb rule for {\it sliding} friction and the force balance conditions on infinitesimal string segments (Figure~\ref{critical}). With increasing $T_2$ the slip arc domain grows and macroscopic failure occurs when it expands to fill the entire contact line.  This is precisely the opposite approach to the global slipping as found for the elastically optimized solutions in Section IV.

\smallskip
The Grashof solution does not minimize the long range part of the elastic energy (Eq.~6). To see this notice that the matching condition between the two tension domains in the Grashof solution inevitably requires a slope discontinuity in the tension profile exactly at the matching point. The influence of this slope discontinuity can be understood by using the Fourier representation in Eqn. 7.  The slope discontinuity produces a tail of nonvanishing amplitudes of the tension profile $t_m$ that persists to very large $m$. Then using Eq.~7 one sees that this increases the macroscopic elastic energy. Interestingly we notice that the Grashof solution {\it does} minimizes a different energy functional that would retain only the short range contribution to the elastic energy in the compliant belt. To see this consider a differential segment of the belt with Young's modulus $Y$ and cross sectional area $A$. If $\delta$ is the amount by which the segment is stretched from its relaxed length $R \, d \phi$ under tension $T$ one finds
\begin{eqnarray}
\frac{T}{A} = Y \frac{\delta}{R \, d \phi} \nonumber
\end{eqnarray}
and a stored elastic energy in the segment
\begin{eqnarray}
dU = \frac{1}{2} \frac{RT^2}{AY} \, d \phi \nonumber
\end{eqnarray}
In this situation the elastic energy functional in Eq.~6 is replaced by a purely {\it local} short range variant
\begin{eqnarray}
U_G = \frac{g_G}{2} \, \int \, T^2 \, d \phi \nonumber
\end{eqnarray}
We see that this energy is minimized by maintaining the {\it minimum possible} tension $T$ profile over the maximum angle of contact (Figure~\ref{critical}) while still maintaining the boundary values $T_1$ and $T_2$. This is precisely the Grashof tension profile shown in Figure~\ref{critical}.  It is important to notice that this construction does not allow any mechanical coupling between neighboring string elements and therefore it is not a model for the compliance of a continuous elastic medium. In essence the Grashof solution replaces the elastic continuum theory by a model with an ensemble of independent {\it local} oscillators.

\smallskip
Alternatively, the Grashof solution can be rationalized as resulting from a transient tension distribution that appears in the belt as it loses contact with the pulley. Since the pulley is being driven from its high tension side one could expect this region to slip before the space of elastic energy-minimizing solutions along the full contact line can be explored.  Were this situation to occur it would actually violate the stability conditions for the capstan in {\it static} equilibrium where the Amontons-Coulomb law predicts instead that the low tension side is more susceptible to slip.   But an even more perplexing feature of this interpretation is that in the Grashof model this one-sided slipping state is predicted to occur even if the load is arbitrarily weak!  Amazingly, despite these pathologies, experimental measurements using force transducers \cite{KMN} and strain gauges \cite{DPT} to probe the tension profile along a belt on a driven pulley find some qualitative features of the Grashof model in the data.  For the {\it driving} pulley, measurements of the tension profile are inverted with respect to the {\it driven} pulley: constant tension is on the high tension side with its value matched to an exponential profile on the low tension side due to slip as the belt contracts. This supports an interpretation of this profile as a transient that is kinetically limited in the regions where the belt exits the pulleys. The inverted Grashof profile cannot be associated with any minimization condition (of which we are aware) for the local elastic energy.  Furthermore, for the entire system, the mechanical energy is not conserved because of the sliding friction on both pulleys and accordingly the rim speed of the driven pulley is less than the rim speed of the driving pulley \cite{Johnson}.

\smallskip

\section{Laboratory Exercises}

\smallskip
It would be both interesting and pedagogically useful as a laboratory project to measure the tension profile in the Atwood's machine to critically test these models, identify whether a strain profile is reproducible or indeterminate, and provide a quantitative basis for further discussion. It is not possible to determine anything about the tension profile on the contact line for an Atwood machine by measuring just the acceleration, since this is determined by the boundary values of $T_1$ and $T_2$ which are the same in all cases.  However, small elastic deformations in the pulley that are induced by the tension profile could be directly visualized and even quantified photoelastically in a darkfield polariscope \cite{Coker}.  In this method a transparent isotropic polymeric solid ({\it e.g.}\ readily-available plexiglass or polyurethane) of constant thickness is sandwiched between left- and right-handed circular polarizers and the intensity pattern of transmitted light is imaged.  If the solid is unstrained, then no light is transmitted.  But if it is strained, then optical birefringence is induced and the polarization of the light changes as it propagates across the material.  Consequently, only a corresponding fraction of the incident light is transmitted through the second polarizer.  In typical experimental settings the polarization state can rotate several times as a light ray passes through a sample.  Therefore the detected intensity is not a single-valued function of local strain and a series of bright-dark stripes is ordinarily observed for spatially-varying strain fields. This method has been applied to measure the local contact forces in quasi-2d granular packings of polymeric disks  \cite{HowellPRL1999, MajumdarNature2005,DanielsRSI2017} .  In particular, Daniels \cite{DanielsRSI2017} gives a tutorial review with detailed methods for how to analyze images and extensive advice on how to choose, machine, and cast suitable photoelastic parts of arbitrary shape.  Commercial systems are also available, where a photoelastic coating is applied to a reflective solid \cite{E.g. micro-measurements.com/photostress}.

We foresee many possibilities for inexpensive custom-built photoelasticity experiments on the Atwood's machine. We have not yet built such an apparatus, but here we propose some promising lines of investigation. The simplest idea is to image the light transmitted through a pulley consisting of a polymeric disk sandwiched between left- and right-circular polarizers with slightly larger radii in order to guide the rope.  Rather than actual rope, it is preferable to use a ribbon or belt of width slightly smaller than the thickness of the polymeric disk so that the strain field is two-dimensional.  Stress will concentrate near the axle, but the bright-dark stripe pattern will exhibit angular variation indicative of the strain field.  Since the pulley is rotationally symmetric, it will suffice to acquire a single image (perhaps of long exposure in order to average over slight manufacturing defects).  But the bright-dark stripe pattern should be visually inspected or imaged versus time to ensure that initial transients damp out and the acceleration becomes constant.  If a video camera is available, a thin annulus of opaque dots or lines could be superposed on one of the polarizers and used to deduce angular acceleration.

\begin{figure}[ht]
\includegraphics[width=\columnwidth]{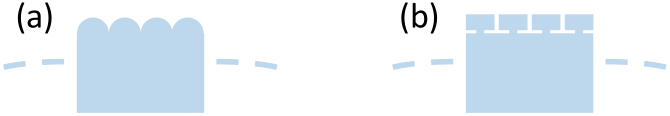}
\caption{Example corrugations along the rim of photoelastic pulleys, designed to enable measurement of local friction and normal forces between pulley and rope.}
\label{pulleys}
\end{figure}

If one eliminates the rotational symmetry of the pulley,  a linear tension profile can no longer be inferred from minimization of the elastic energy (Figure~\ref{pulleys}).  This could be arranged by machining large holes in the photoelastic disk, with shapes that are either circular or to in effect make spokes. Another variation would be to use a pulley that is rigid except for an outer rim of compliant photoelastic polymer with width comparable to the disk thickness and small compared to it radius.  This situation is equivalent to a rigid pulley with a compliant rope since elastic energy is stored only along the rim.

\section{Conclusion}

\smallskip
In the presentation of the massive Atwood's machine in an introductory or intermediate course in mechanics the distribution of string tensions along the contact line is unknown and unknowable.  For  the discerning student this invites the very relevant questions of ``what is the actual tension profile" or at least ``how is it determined". This paper compares two plausible but very different resolutions of this problem. Let us summarize the main results.

\smallskip
(1) When the no slip condition is applied to a unstretchable string and perfectly rigid pulley the tension and friction profiles are indeterminate. As the students are instructed in this case only the boundary values of the tension $T_1$ and $T_2$ are specified by the equation for rigid body rotation. Coulomb friction does not allow one to further determine the force distribution along the contact line and instead there are a large family of admissible solutions (Section III).

\smallskip
(2) If one enforces a no slip condition on the contact line and allows for the elastic compliance of the string and/or pulley there is a well defined solution for the tension profile that minimizes the stored elastic energy. This solution linearly interpolates between the boundary values as one might have naively supposed. However, above a critical load the elastically optimized solution is instead a piecewise continuous solution containing two domains. One domain (on the low tension side) self organizes to the critical state on the verge of slipping with exponentially growing friction.  This solution is matched to the value and slope of a second linear tension profile in which the distributed friction is constant (Section IV).

\smallskip
(3) If one posits that slipping motion always occurs on the exit side of the pulley one arrives at the complementary piecewise continuous solution for arbitrary loading.  In this solution (known as the Grashof model) one low tension domain has constant tension with no friction (it floats on the contact line) matched to the value but necessarily {\it not} the slope of the profile in a second domain with an exponential friction/tension distribution (Section V).

\smallskip
The physical resolution could be further complicated by history dependence, slip and creep along the contact. All these scenarios offer interesting possibilities for the development of an advanced laboratory project to test them.

\section*{Acknowledgement}

We thank Rob Carpick and Mark Robbins for useful discussions and correspondence on this topic.

\end{document}